# Is A Quantum Stabilizer Code Degenerate or Nondegenerate for Pauli Channel?

Fangying Xiao, Hanwu Chen

*Abstract*—Mapping an error syndrome to the error operator is the core of quantum decoding network and is also the key step of recovery. The definitions of the bit-flip error syndrome matrix and the phase-flip error syndrome matrix were presented, and then the error syndromes of quantum errors were expressed in terms of the columns of the bit-flip error syndrome matrix and the phase-flip error syndrome matrix. It also showed that the error syndrome matrices of a stabilizer code are determined by its check matrix, which is similar to the classical case. So, the error-detection and recovery techniques of classical linear codes can be applied to quantum stabilizer codes after some modifications. Some necessary and/or sufficient conditions for the stabilizer code over GF(2) is degenerate or nondegenerate for Pauli channel based on the relationship between the error syndrome matrices and the check matrix was presented. A new way to find the minimum distance of the quantum stabilizer codes based on their check matrices was presented, and followed from which we proved that the performance of degenerate quantum code outperform (at least have the same performance) nondegenerate quantum code for Pauli channel .

## I. Introduction

Quantum information can be protected by encoding it into a quantum error-correcting code. Quantum error-correcting codes are quite similar to classical codes in many respects, for example, an error is identified by measuring the error syndrome, and then corrected as appropriate, just as in the classical case. However, there is an interesting class of quantum codes known as degenerate codes [1],[2],[3],[4] possessing a string property unknown in classical codes. For classical error-correcting codes errors on different bits necessarily lead to different corrupted codewords. But for a degenerate quantum code, the error syndrome is not unique, and error syndromes are only repeated when $E^{\dagger}F$ belongs to the stabilizer $S$, implying that $E$ and $F$ act the same way on the codewords. So they can sometimes be used to correct more errors than they can identify. The phenomenon of degenerate quantum codes is a sort of good news-bad news situation for quantum codes. The bad news is that some of the proof techniques used classically to prove bounds on error-correction fall down because they can't be applied to degenerate quantum codes, for example, the quantum Hamming bound. The good news is that degenerate quantum codes seem to be among the most interesting quantum codes. They are known to outperform all nondegenerate quantum codes for very noisy quantum channel (for example, Pauli channel)[5],[6],[7] and have important applications in purifying quantum states[8] and proving the security of quantum communication protocols[9],]10]. It is possible that they are able to store quantum information more efficiently than any nondegenerate code, because distinct errors do not necessarily have to take the code space to orthogonal space.

The decoding network of quantum code contains three parts, namely, error detection, error correction, decoding. Mapping error syndromes to error operators is the core of quantum decoding network and is also the key step to realize quantum error correction. In the third section, we presented the definitions of the bit-flip error syndrome matrix and the phase-flip error syndrome matrix, and then expressed the error syndromes of all quantum errors in terms of the columns of the bit-flip error syndrome matrix and the phase-flip error syndrome matrix. In the fourth section, it showed that the error syndrome matrices of a stabilizer code are determined by its check matrix, which is similar to the classical case. So, the error detection and correction techniques of classical linear codes can be applied to quantum stabilizer codes after some modifications.

Until now, no technique has been developed to determine whether or not a stabilizer code is degenerate or nondegenerate other than exhaustive search. In the fifth section, based on the relationship between the error syndrome matrices and the check matrix of quantum stabilizer code some necessary and/or sufficient conditions for the stabilizer code over GF(2) is degenerate or nondegenerate were presented for Pauli channel. A new way to find the minimum distance of quantum stabilizer codes based on their check matrices was also presented, and followed from which we proved that the performance of degenerate quantum code are outperform (at least have the same performance as) nondegenerate quantum code for Pauli channel. We hope that these results will helpful for design quantum degenerate codes [11],[12] with good performance.

## II. Preliminary

The commutator and anti-commutator[1] of the operator $A$ and the operator $B$ are $[A,B] = AB-BA$ and $\{A, B\} = AB+BA$ respectively. If $[A, B] = 0$, then the operator $A$ and the operator $B$ are said to *commute*. If $\{A, B\} = 0$, then $A$ and $B$ are said to *anticommute*.

Four extremely useful matrices called *Pauli matrices* with their corresponding notations are described following:

$$I = \begin{bmatrix} 1 & 0 \\ 0 & 1 \end{bmatrix}, X = \begin{bmatrix} 0 & 1 \\ 1 & 0 \end{bmatrix}, Z = \begin{bmatrix} 1 & 0 \\ 0 & -1 \end{bmatrix}, Y = \begin{bmatrix} 0 & -i \\ i & 0 \end{bmatrix}$$

from which we see that $[X, Y] = 2iZ$, $[Y, Z] = 2iX$, $[Z, X] = 2iY$.

The Pauli group $G_1$ on 1 qubit is the matrix group consisting all of the Pauli matrices $I$, $X$, $Y$, $Z$, together with multiplicative factors $\pm 1, \pm i$, i.e.,

$$G_1 = \{\pm I, \pm iI, \pm X, \pm iX, \pm Y, \pm iY, \pm Z, \pm iZ\}$$

The Pauli group $G_n$ on $n$-qubit is the group generated by the operators described above applied to each of $n$ qubits in the tensor product Hilbert space $C^{2^n}$,

Xiao Fangying and Chen Hanwu are with the School of Computer Science and Engineering, Southeast University, Nanjing, Jiangsu 210096, China

$$G_n = \left\{ g \,\middle|\, \begin{array}{l} g = i^c g_1 \otimes g_2 \otimes \cdots \otimes g_n, \\ g_j \in \{I, X, Z, Y\}, c \in \{0,1,2,3\} \end{array} \right\},$$

where $i = \sqrt{-1}$.

It follows from the definition of the Pauli group that if $A, B \in G_n$, then $[A, B] = 0$ or $\{A, B\} = 0$. Since $Y = iZX$, so any element of $G_n$ can be denoted as $g = i^c \bigotimes_{j=1}^{n} \left( X^{a_j} \cdot Z^{b_j} \right)$, where $a_i, b_j \in F_2$.

A homomorphism from $G_n$ onto $F_2^{2n}$ defined as

$$\varphi : G_n \mapsto F_2^{2n} \qquad (1)$$

that is, $\varphi(g) = (\alpha | \beta) \triangleq (\varphi(g)_X | \varphi(g)_Z)$, where

$$g = i^c \bigotimes_{j=1}^{n} \left( X^{a_j} \cdot Z^{b_j} \right) \in G_n, \quad c \in \{0,1,2,3\},$$

$\alpha = (a_1, a_2, \cdots, a_n)$ and $\beta = (b_1, b_2, \cdots, b_n) \in F_2^n$.

It follows from the above definition that the Pauli matrices $I, X, Y, Z$ can be mapped to the following binary vectors:

$\varphi(I) = (0 | 0), \varphi(X) = (1 | 0), \varphi(Z) = (0 | 1), \varphi(Y) = (1 | 1)$.

An isomorphism from $F_2^{2n}$ onto $\overline{G}_n$ define as:

$$\varphi' : F_2^{2n} \mapsto \overline{G}_n, \overline{G}_n = G_n \big/ \{\pm 1, \pm i\} \qquad (2)$$

that is, $\varphi'((\alpha|\beta)) = \bigotimes_{j=1}^{n} \left( X^{a_j} \cdot Z^{b_j} \right)$.

For example, $\varphi(iXIZYXZY) = (1001101 | 0011011)$ and $\varphi'(1001101 | 0011011) = XIZYXZY$.

Let $g \in G_n$, such that $\varphi(g) = (\varphi(g)_X | \varphi(g)_Z)$, then the *quantum weight* of $g$, written $w_Q(g)$, is the number of components that is not $I$, that is

$$w_Q(g) = w(\varphi(g)_X) + w(\varphi(g)_Z) - w(\varphi(g)_X \cdot \varphi(g)_Z)$$

where $w(u)$ is the Hamming weight of $u$ and "." is the inner product.

The *symplectic weight* of a vector $v = (v_1 | v_2)$, where $v_1, v_2 \in F_2^n$, written $w_s(v)$, is

$$w_s(v) = w(v_1) + w(v_2) - w(v_1 \cdot v_2).$$

As a result, $w_Q(g) = w_s(\varphi(g))$.

Let $\{M_1, M_2, \cdots, M_{n-k}\}$ are the generators of the stabilizer $S$, where $M_i \in G_n$ $(1 \leq i \leq n-k)$, $H$ is an $(n-k) \times 2n$ matrix over GF(2) whose rows contains the vectors $\varphi(M_1), \varphi(M_2), \cdots, \varphi(M_{n-k})$, that is,

$$H = \begin{bmatrix} \varphi(M_1) \\ \vdots \\ \varphi(M_{n-k}) \end{bmatrix} \triangleq [H_X | H_Z] \qquad (3)$$

$H$ is called the *check matrix*[1] of $S$.

## III. ERROR SYNDROME

Suppose $S = \langle M_1, M_2, \cdots, M_{n-k} \rangle$ is the stabilizer of a quantum stabilizer code $C(S)$. Let the encoded state $|\phi\rangle$ was suffered from some inference denoted $E \in G_n$ during transmitting on a noisy channel which transmitted the encoded state to $|\phi'\rangle = E|\phi\rangle$ (say). The error detection can be done by simply measure the generators of the stabilizer, as shown in Figure 1 [1]. It will give us a list of eigenvalues, the error syndrome, which will tell us whether the error $E$ commutes or anticommutes with each of the generators.

From figure 1 we see that if $[E, M_i] = 0$, then $s_i = 0$; if $\{E, M_i\} = 0$, then $s_i = 1$.

The vector $s_E = \left( s_E^1, s_E^2, \cdots, s_E^{n-k} \right) \in F_2^{n-k}$ represents the

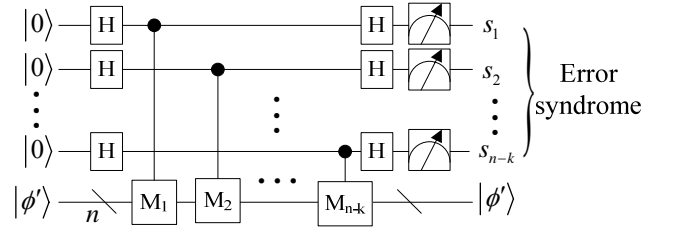

Fig. 1. Syndrome measurement.

error syndrome of the error $E \in G_n$, where if $[E, M_j] = 0$, then $s_E^j = 0$; if $\{E, M_j\} = 0$, then $s_E^j = 1$ where, $j = 1, 2, \cdots, n-k$.

The operator $I^{\otimes(i-1)} \otimes X \otimes I^{\otimes(n-i)}$ denotes the bit-flip error on the $i$-th qubit, written $X_i$, i.e., $X_i = I^{\otimes(i-1)} \otimes X \otimes I^{\otimes(n-i)}$, then the error syndrome of $X_i$ is $s_{X_i} = \left( s_{X_i}^1, s_{X_i}^2, \cdots, s_{X_i}^{n-k} \right) \in F_2^{n-k}$.

Set the rows of $s_X$ to be $s_{X_1}, s_{X_2}, \cdots, s_{X_n}$, then the $n \times (n-k)$ matrix

$$s_X = \begin{bmatrix} s_{X_1} \\ s_{X_2} \\ \vdots \\ s_{X_n} \end{bmatrix} = \begin{bmatrix} s_{X_1}^1 & s_{X_1}^2 & \cdots & s_{X_1}^{n-k} \\ s_{X_2}^1 & s_{X_2}^2 & \cdots & s_{X_2}^{n-k} \\ \vdots & \vdots & \cdots & \vdots \\ s_{X_n}^1 & s_{X_n}^2 & \cdots & s_{X_n}^{n-k} \end{bmatrix} \in F_2^{n \times (n-k)}$$

is called the *bit-flip error syndrome matrix*(BSM).

More generally, if the bit-flip errors happen to multiple qubits, written $E_X = i^{c_X} \prod_{v=1}^{n} X_v^{a_v}$, where $c_X \in \{0,1,2,3\}$, $a_v \in F_2$. Suppose $a = [a_1, a_2, \cdots, a_n] \in F_2^n$, if $a_v = 1$, then a bit-flip error happens to the $v$-th qubit; if $a_v = 0$, then no error happens to the $v$-th qubit. Since

$$E_X \cdot M_j = i^{c_X} \prod_{v=1}^{n} X_v^{a_v} \cdot M_j$$

$$= (-1)^{(a_n \cdot s_{X_n}^j)} i^{c_X} \prod_{v=1}^{n-1} X_v^{a_v} \cdot M_j \cdot X_n^{a_n}$$

$$= (-1)^{\sum_{u=1}^{n}(a_u \cdot s_{X_u}^j)} M_j \cdot i^{c_X} \prod_{v=1}^{n} X_v^{a_v}$$

$$= (-1)^{\sum_{u=1}^{n}(a_u \cdot s_{X_u}^j)} M_j \cdot E_X$$

Let $s_X^j = \left[s_{X_1}^j, s_{X_2}^j, \cdots, s_{X_n}^j\right]^T \in F_2^n$, then

$$s_{E_X}^j = \sum_{u=1}^{n}\left(a_u \cdot s_{X_u}^j\right) = a \cdot s_X^j$$

Therefore, the error syndrome of $E_X = i^{c_X} \prod_{v=1}^{n} X_v^{a_v}$ is

$$s_{E_X} = \left[a \cdot s_X^1, a \cdot s_X^2, \cdots, a \cdot s_X^{n-k}\right] = a \cdot s_X \quad (4)$$

It follows from Eq (4) that the error syndrome $s_{E_X}$ of any error $E_X = i^{c_X} \prod_{v=1}^{n} X_v^{a_v}$ can be expressed as a linear combination of the error syndrome $s_{X_v}$ of $X_v (v=1,2,\cdots,n)$.

The operator $Z_i = I^{\otimes(i-1)} \otimes Z \otimes I^{\otimes(n-i)} (1 \le i \le n)$ denotes the phase-flip error on the $i$-th qubit. Similarly, the matrix

$$s_Z = \begin{bmatrix} s_{Z_1}^1 & s_{Z_1}^2 & \cdots & s_{Z_1}^{n-k} \\ s_{Z_2}^1 & s_{Z_2}^2 & \cdots & s_{Z_2}^{n-k} \\ \vdots & \vdots & \cdots & \vdots \\ s_{Z_n}^1 & s_{Z_n}^2 & \cdots & s_{Z_n}^{n-k} \end{bmatrix} \in F_2^{n \times (n-k)}$$

where if $\left[Z_i, M_j\right] = 0$ then $s_{Z_i}^j = 0$ and if $\{Z_i, M_j\} = 0$ then $s_{Z_i}^j = 1$, $i=1,2,\cdots,n$, $j=1,2,\cdots,n-k$, is called the *phase-flip error syndrome matrix* (PSM).

If the phase-flip errors happen to multiple qubits, that is $E_Z = i^{c_Z} \prod_{l=1}^{n} Z_l^{b_l}$, where $b_l \in F_2, c_Z \in \{0,1,2,3\}$.

Let $b = [b_1, b_2, \cdots, b_n] \in F_2^n$, Similarly, the error syndrome of $E_Z = i^{c_Z} \prod_{l=1}^{n} Z_l^{b_l}$ is

$$s_{E_Z} = \left[b \cdot s_Z^1, b \cdot s_Z^2, \cdots, b \cdot s_Z^{n-k}\right] = b \cdot s_Z$$

From which we see that the error syndrome $s_{E_Z}$ of any error $E_Z = i^{c_Z} \prod_{l=1}^{n} Z_l^{b_l}$ can be expressed as a linear combination of the error syndrome $s_{Z_l}$ of $Z_l (l=1,2,\cdots,n)$.

Suppose $Y_m = I^{\otimes(m-1)} \otimes Y \otimes I^{\otimes(n-m)} (1 \le m \le n)$ which is the case of that two kinds of errors (bit-flip and phase-flip) happened contemporarily on the $m$-th qubit. Since $Y=iZX$, so

$$Y_m = iZ_m \cdot X_m \quad (5)$$

It follows from Eq (5) that an arbitrary error $E_P \in G_n$ can be expressed as

$$E_P = i^{c_P} \prod_{v=1}^{n} X_v^{a_v} \cdot \prod_{l=1}^{n} Z_l^{b_l}$$

where $a_v, b_l \in F_2, c_P \in \{0,1,2,3\}$. Let $a = [a_1, a_2, \cdots, a_n]$ and $b = [b_1, b_2, \cdots, b_n] \in F_2^n$, such that

$$E_P \cdot M_j = i^{c_P} \prod_{v=1}^{n} X_v^{a_v} \cdot \prod_{l=1}^{n} Z_l^{b_l} \cdot M_j$$

$$= (-1)^{b \cdot s_Z^j} i^{c_P} \prod_{v=1}^{n} X_v^{a_v} \cdot M_j \cdot \prod_{l=1}^{n} Z_l^{b_l}$$

$$= (-1)^{b \cdot s_Z^j + a \cdot s_X^j} M_j \cdot \left( i^{c_P} \prod_{v=1}^{n} X_v^{a_v} \cdot \prod_{l=1}^{n} Z_l^{b_l} \right)$$

$$= (-1)^{b \cdot s_Z^j + a \cdot s_X^j} M_j \cdot E_P$$

Then

$$s_{E_P}^j = a \cdot s_X^j + b \cdot s_Z^j$$

Therefore, the error syndrome of $E_P = i^{c_P} \prod_{v=1}^{n} X_v^{a_v} \cdot \prod_{l=1}^{n} Z_l^{b_l}$ is

$$s_{E_P} = \left(a \cdot s_X^1 + b \cdot s_Z^1, a \cdot s_X^2 + b \cdot s_Z^2, \cdots, a \cdot s_X^{n-k} + b \cdot s_Z^{n-k}\right)$$
$$= \left(a \cdot s_X^1, a \cdot s_X^2, \cdots, a \cdot s_X^{n-k}\right) + \left(b \cdot s_Z^1, b \cdot s_Z^2, \cdots, b \cdot s_Z^{n-k}\right) \quad (6)$$
$$= a \cdot s_X + b \cdot s_Z$$

It follows from Eq (6) that the error syndrome $s_{E_P}$ of any error $E_P \in G_n$ can be expressed as a linear combination of the error syndrome $s_{Z_l}$ of $Z_l(l=1,2,\cdots,n)$ and the error syndrome $s_{X_v}$ of $X_v(v=1,2,\cdots,n)$.

Remarkably, suitable error syndrome measurements would collapse an arbitrary error (including coherent superpositions of bit-flip and phase-flip errors) into the discrete set of only bit-flip and/or phase-flip errors, because it can be expressed as a superposition of basis operations—the error basis (which is here given by the Pauli matrices). And these discrete Pauli errors can be easily reversed to recover the original state. So, it only has to find out the error syndromes of the bit-flip error $X_v(1 \le v \le n)$ and the phase-flip error $Z_l(1 \le l \le n)$ on 1-qubit, and then the error syndrome of an arbitrary error can be expressed in terms of the columns of the bit-flip error syndrome matrix and the phase-flip error syndrome matrix according to Eq (6).

From previous analysis we knew that the errors in the set $\{\pm E_s, \pm iE_s\}$ can be mapped to the same error syndrome $s \in F_2^{n-k}$, since from an observational point of view the quantum states corrupted by the errors with different global phase factor are identical. For this reason we may ignore the global phase factors as being irrelevant to the observed properties of the physical system and having no affect to the recovery. In the rest of the article we will only consider the errors in $\overline{G}_n$.

## IV. ANALYZE THE ERROR SYNDROME OF STABILIZER CODES

Suppose $C(S)$ is a stabilizer code with stabilizer $S = \langle M_1, M_2, \cdots, M_{n-k} \rangle$ whose check matrix is $H_{(n-k) \times 2n} = [H_X | H_Z]$. Whether a bit-flip error $X_i$ commutes with or anticommutes with $M_j$ dependent on whether or not the $i$-th component of $M_j$ is $Z$ or $Y$. If $M_{ji} = Z$ or $Y$, then $\{X_i, M_j\} = 0$, otherwise $[X_i, M_j] = 0$. It follows from Eq (3) that the syndrome of the bit-flip error in the $i$-th qubit, $X_i (1 \leq i \leq n)$, corresponds to the $n+i$-th column of $H$ which is also the $i$-th column of the sub-matrix $H_Z$. Similarly, the syndrome of the phase-flip error in the $i$-th qubit, $Z_i (1 \leq i \leq n)$, corresponds to the $i$-th column of $H$ which is also the $i$-th column of the sub-matrix $H_X$. That is,

$$X_i M_j = (-1)^{\varphi(M_j)_{n+i}} M_j X_i = (-1)^{H_{j,n+i}} M_j X_i \qquad (7)$$

$$Z_i M_j = (-1)^{\varphi(M_j)_i} M_j Z_i = (-1)^{H_{ji}} M_j Z_i \qquad (8)$$

where $\varphi(M_j)_i$ is the $i$-th component of the $2n$-dimentional binary vector $\varphi(M_j)$ and $H_{ji}$ is the element at row $j$, column $i$ of matrix $H$, $1 \leq j \leq n-k, 1 \leq i \leq 2n$. It follows from (7) and (8) that, for a stabilizer code $C(S)$ the error syndromes of $X_i$ and $Z_i$ are

$$s_{X_i} = (H_{1,n+i}, H_{2,n+i}, \cdots, H_{n-k,n+i})$$
$$s_{Z_i} = (H_{1i}, H_{2i}, \cdots, H_{n-k,i})$$

where $s_{X_i}^j = H_{j,n+i}$ and $s_{Z_i}^j = H_{ji}$.

Let
$$H = (H_X | H_Z) = (H_X^1 \ H_X^2 \ \cdots \ H_X^n | H_Z^1 \ H_Z^2 \ \cdots \ H_Z^n)$$

where $H_X^i (1 \leq i \leq n)$ is the $i$-th column of $H_X$ and $H_Z^i (1 \leq i \leq n)$ is the $i$-th column of $H_Z$. Then

$$s_{X_i} = \left(H_Z^i\right)^T, s_{Z_i} = \left(H_X^i\right)^T$$

The following theorem now follows easily after the discussion above.

*Theorem 1.* If the check matrix of a stabilizer code $C(S)$ is $(H_X | H_Z)$, then $s_X = H_Z^T$ and $s_Z = H_X^T$.

Therefore, the error syndrome of an arbitrary error $E \in \overline{G}_n$ can be expressed in terms of the columns of the check matrix, i.e.,

$$\begin{aligned} s_E &= \varphi(E)\left(\Lambda\left[H_Z | H_X\right]\right)^T \\ &= [\varphi(E)_X | \varphi(E)_Z]\left([H_Z | H_X]\right)^T \\ &= \varphi(E)_X H_Z^T + \varphi(E)_Z H_X^T \end{aligned}$$

where $\Lambda = \begin{bmatrix} 0 & I_n \\ I_n & 0 \end{bmatrix}$ and $\varphi(E) = [\varphi(E)_X | \varphi(E)_Z]$.

It has come to light that the error-detection and recovery of the classical codes are totally dependent on their parity check matrices. Theorem 1 makes a connection between the error syndrome matrices (BSM and PSM) and the check matrix of the stabilizer code. Thereby, expressing the error syndrome as a linearly combination of the columns of BSM and PSM is transformed into expressing it as a linearly combination of the columns of the check matrix. So, the error-detection and recovery of the quantum stabilizer codes are also totally dependent on their check matrices, which is similar to the classical case. Therefore, the error-detection and recovery schemes and techniques of the classical codes can be applied to quantum stabilizer codes easily.

## V. IS A STABILIZER CODE DEGENERATE OR NONDEGENERATE?

In this section we describe some necessary and/or sufficient conditions for a stabilizer code over GF(2) is degenerate or nondegenerate based on its check matrix for Pauli channel which can also be applied to depolarization channel.

Strictly speaking, degeneracy is not a property of a quantum code alone, but a property of a code together with a family of errors it is designed to correct [5]. Here we will show a definition of degenerate and nondegenerate stabilizer codes for Pauli channel.

Let $\overline{G}_{n|t} = \{g \in \overline{G}_n \mid w_Q(g) \leq t\}$, consider a mapping from $\overline{G}_{n|t}$ to $F_2^{2n}$:

$$f : \overline{G}_{n|t} \mapsto F_2^{n-k}$$

that is, for $\forall g \in \overline{G}_{n|t}$, $f(g) = (v_1, v_2, \cdots, v_{n-k}) \in F_2^{n-k}$. If $[g, M_i] = 0$ then $v_i = 0$, if $\{g, M_i\} = 0$ then $v_i = 1$. If $f$ is nonsingular, we say that the stabilizer codes with parameters $[\![n, k, 2t+1]\!]$ and $[\![n, k, 2t+2]\!]$ are *nondegenerate*, otherwise, they are *degenerate*.

In the rest of this section we only discuss the stabilizer codes with parameters $[\![n, k, 2t+1]\!]$, and all the conclusions can also be applied to stabilizer codes with parameters $[\![n, k, 2t+1]\!]$.

*Lemma 1.* If the set of vectors $v = \{a_1, a_2, \cdots, a_{2m}\}$ is linearly independent, then the sum of any set of $m$ vectors of $v$ are different from each other.

*Proof*: suppose the vectors $a_{i_1}, a_{i_2}, \cdots, a_{i_m}$ and the vectors $a_{j_1}, a_{j_2}, \cdots, a_{j_m}$, where

$\{i_1, i_2, \cdots, i_m\} \cap \{j_1, j_2, \cdots, j_m\} < m$, satisfy

$$\sum_{k=1}^{m} c_k a_{i_k} = \sum_{l=1}^{m} c'_l a_{j_l}, \text{ with } c_p = 1 \text{ and } c_k, c'_l \in \{0,1\}$$

Then, we have $\sum_{k=1}^{m} c_k a_{i_k} + \sum_{l=1}^{m} c'_l b_{j_l} = 0$. Thus $v$ is fails to be linearly independent. Therefore, we produce a contradiction.
□

*Theorem 2.* Let $[H_X | H_Z]_{n \times 2(n-k)}$ be a check matrix such that any $4t$ columns linearly independent, then the stabilizer code $C(S)=[[n,k,2t+1]]$ with check matrix $[H_X | H_Z]$ is nondegenerate.

*Proof:* Since $C(S)$ is nondegenerate if (and only if) for arbitrary $E_1, E_2 \in \overline{G}_{n|t}$ and $E_1 \neq E_2$, they satisfy $s_{E_1} \neq s_{E_2}$. We will perform two steps in order to prove the theorem.

(i) Let us assume as before that

$$(H_X | H_Z) = (H_X^1 \; H_X^2 \; \cdots \; H_X^n | H_Z^1 \; H_Z^2 \; \cdots \; H_Z^n),$$

assume further that $v_{m,p}$ is the sum of the $i_1$th, $i_2$th, $\cdots, i_m$th columns of $H_X$ and the $j_1$th, $j_2$th, $\cdots, j_p$th columns of $H_Z$. That is $v_{m,p} = \sum_{a=1}^{m} H_X^{i_a} + \sum_{b=1}^{p} H_Z^{j_b} \in F_2^{n-k}$. Since any $4t$ columns of $[H_X | H_Z]$ are linearly independent, it follows from lemma 1 that if $m$ and $p$ satisfy $1 \leq m + p \leq 2t$, then we will get different $v_{m,p}$ with respect to different sets $\{i_1, i_2, \cdots, i_m\}$ and $\{j_1, j_2, \cdots, j_p\}$.

(ii) Let

$$u = (u_1 | u_2) = (\{j_1, j_2, \cdots, j_p\} | \{i_1, i_2, \cdots, i_m\}) \in F_2^{2n},$$

where $u_1$ and $u_2 \in F_2^n$, $\{i_1, i_2, \cdots, i_m\}$ and $\{j_1, j_2, \cdots, j_p\}$ are supports of $u_1$ and $u_2$ respectively. Therefore

$$v_{m,p} = \sum_{a=1}^{m} H_X^{i_a} + \sum_{b=1}^{p} H_Z^{j_b} = \left(u(H_Z | H_X)^T\right)^T$$
$$= \left(u_1 H_Z^T + u_2 H_X^T\right)^T \in F_2^{(n-k)}$$

It follows from Eq (2) that, the operator corresponding to $u$ is

$$E = \varphi'(u) = \prod_{k=1}^{n} Z_k^{u_{2k}} \cdot \prod_{l=1}^{n} X_l^{u_{1l}} = \prod_{a=1}^{m} Z_{i_a} \cdot \prod_{b=1}^{p} X_{j_b}$$

So, the error syndrome of $E$, $s_E$, is $(v_{m,p})^T$.

If the sets $\{i_1, i_2, \cdots, i_m\}$ and $\{j_1, j_2, \cdots, j_n\}$ satisfy the following conditions:

$$0 \leq m, p \leq t, 1 \leq m + p \leq 2t \text{ and}$$
$$|\{i_1, i_2, \cdots, i_m\} \cup \{j_1, j_2, \cdots, j_p\}| \leq t$$

then, $0 \leq w(u_1), w(u_2) \leq t$ and $1 \leq w(u_1 \cdot u_2) \leq t$. Therefore, $w_Q(E) = w(u_1) + w(u_2) - w(u_1 \cdot u_2) \leq t$.

It follows from Eq (2) that, if we pick different columns from $[H_X | H_Z]$ which corresponds to different sets $\{i_1, i_2, \cdots, i_m\}$ and $\{j_1, j_2, \cdots, j_p\}$, then they will correspond to different error $E \in \overline{G}_{n|t}$.

Finally, combining (i) and (ii) we get that the stabilizer code $C(S)$ is nondegenerate.
□

Note that the reverse of theorem 2 is generally not true. That is, there would exist a set of $4t$ linearly dependent columns of $[H_X | H_Z]_{n \times 2(n-k)}$ even if the stabilizer code with parameters $[[n,k,2t+1]]$ is nondegenerate. Because if the $4t$ linearly dependent columns satisfy the following conditions:

$0 \leq m, p \leq 4t$, $|\{i_1, i_2, \cdots, i_m\}| + |\{j_1, j_2, \cdots, j_p\}| = 4t$ and $t < |\{i_1, \cdots, i_m\} \cap \{j_1, \cdots, j_p\}| < 2t$.

Then, at least one of the two corresponding errors that have the same error syndrome has quantum weight lager than $t$, which does not violate the definition of nondegenerate code.

We have the following corollary.

*Corollary 1.* if the stabilizer code $C(S)=[[n,k,2t+1]]$ is nondegenerate, then any $2t$ columns of its check matrix $[H_X | H_Z]$ are linearly independent.

*Proof:* Suppose the $i_1$th, $i_2$th, $\cdots, i_m$th columns of $H_X$ and the $j_1$th, $j_2$th, $\cdots, j_p$th columns of $H_Z$ are linearly dependent, where $i_1, i_2, \cdots, i_m$ and $j_1, j_2, \cdots, j_p \in \{1, 2, \cdots, n\}$ and they satisfy the conditions $0 \leq m, p \leq 2t$ and $|\{i_1, i_2, \cdots, i_m\}| + |\{j_1, j_2, \cdots, j_p\}| = 2t$. Then

$$v_{m,p} = \sum_{a=1}^{m} H_X^{i_a} + \sum_{b=1}^{p} H_Z^{j_b} = 0 \quad (9)$$

Pick a subset $\{i_{k_1}, i_{k_2}, \cdots, i_{k_a}\}$ from $\{i_1, i_2, \cdots, i_m\}$ with $a$ elements and a subset $\{j_{l_1}, j_{l_2}, \cdots, j_{l_b}\}$ from $\{j_1, j_2, \cdots, j_p\}$ with $b$ elements, where $a+b=t$. Let

$$E_{a,b} = \prod_{c=1}^{b} X_{j_{l_c}} \cdot \prod_{d=1}^{a} Z_{i_{k_d}}$$
$$E_{m-a,p-b} = \prod_{c \in \{j_1, j_2, \cdots, j_p\} - \{j_{l_1}, j_{l_2}, \cdots, j_{l_b}\}} X_c \cdot \prod_{d \in \{i_1, i_2, \cdots, i_m\} - \{i_{k_1}, i_{k_2}, \cdots, i_{k_a}\}} Z_d$$

Then, $w_Q(E_{a,b}) \leq t, w_Q(E_{m-a,p-b}) \leq t$

It follows from Eq (9) that, $E_{a,b}$ and $E_{m-a,p-b}$ possess the same error syndrome. Therefore, we produce a contradiction.
□

The theorem 2 and corollary 1 are the sufficient condition and necessary condition for stabilizer codes to be nondegenerate respectively. We have the following theorem that describes a necessary and sufficient condition for a stabilizer code to be nondegenerate.

*Theorem 3.* A stabilizer code $C=[[n,k,2t+1]]$ is nondegenernate if and only if the number of elements in the set

$$\left\{\left(\sum_{a=1}^{m}H_X^{i_a}+\sum_{b=1}^{p}H_Z^{j_b}\right)\left|\begin{array}{l}0\leq m\leq t, 0\leq p\leq t\\ i_1,i_2,\cdots,i_m,j_1,j_2,\cdots,j_p\in\{1,2,\cdots,n\}\\ 0<\left|\{i_1,i_2,\cdots,i_m\}\cup\{j_1,j_2,\cdots,j_p\}\right|\leq t\end{array}\right.\right\}$$

is $\sum_{i=1}^{t}\binom{n}{i}3^i$ or $\sum_{j=0}^{t}\left(\binom{n}{j}\cdot\sum_{l=1}^{t-j}\binom{2n-2j}{l}\right)$.

*Proof*: (a) For Pauli channel, a stabilizer code $C$ is nondegenerate if (and only if) the different errors in its correctible error set $\overline{G}_{n|t}-\{I\}$ are mapped to different error syndromes. Since the number of elements in $\overline{G}_{n|t}-\{I\}$ is $\left|\overline{G}_{n|t}-\{I\}\right|=\sum_{i=1}^{t}\binom{n}{i}3^i$.

Let $E$ is an arbitrary operator of $\overline{G}_{n|t}-\{I\}$ and suppose $\varphi(E)=(u|v)=(u_1u_2\cdots u_n|v_1v_2\cdots v_n)$ with $u_{i_1}=u_{i_2}=\cdots=u_{i_m}=1$ and $v_{j_1}=v_{j_2}=\cdots=v_{j_p}=1$, where $i_1,\cdots,i_m,j_1,\cdots,j_p\in\{1,2,\cdots,n\}$. Then, $i_1,i_2,\cdots,i_m$ and $j_1,j_2,\cdots,j_p$ satisfy the following conditions

$0\leq m,p\leq t$, $0<\left|\{i_1,i_2,\cdots,i_m\}\cup\{j_1,j_2,\cdots,j_p\}\right|\leq t$.

Thus

$$\overline{G}_{n|t}-\{I\}=\left\{\varphi'((u|v))\left|\begin{array}{l}0\leq m\leq t, 0\leq p\leq t\\ i_1,\cdots,i_m,j_1,\cdots,j_p\in\{1,2,\cdots,n\}\\ (u|v)=(\{j_1,\cdots,j_p\}|\{i_1,\cdots,i_m\})\\ 1\leq w_s((u|v))\leq t\end{array}\right.\right\}$$

The error syndrome of $E$ is $s_E=uH_Z^T+vH_X^T$. Then, the error syndromes of the errors in $\overline{G}_{n|t}-\{I\}$ are

$$A=\left\{\left(\sum_{a=1}^{m}H_X^{i_a}+\sum_{b=1}^{p}H_Z^{j_b}\right)^T\left|\begin{array}{l}0\leq m\leq t, 0\leq p\leq t\\ i_1,i_2,\cdots,i_m,j_1,j_2,\cdots,j_p\in\{1,2,\cdots,n\}\\ 0<\left|\{i_1,i_2,\cdots,i_m\}\cup\{j_1,j_2,\cdots,j_p\}\right|\leq t\end{array}\right.\right\}$$

The number of elements in $A$ satisfies

$$n_A\leq\sum_{j=0}^{t}\left(\binom{n}{j}\cdot\sum_{l=1}^{t-j}\binom{2n-2j}{l}\right) \quad (10)$$

If the different errors in $\overline{G}_{n|t}-\{I\}$ are mapped to different error syndromes, then

$$\left|\overline{G}_{n|t}-\{I\}\right|=n_A=\sum_{j=0}^{t}\left(\binom{n}{j}\cdot\sum_{l=1}^{t-j}\binom{2n-2j}{l}\right)$$

(b) Conversely, it follows from Eq (10) that, if the number of elements in $A$ is $\sum_{j=0}^{t}\left(\binom{n}{j}\cdot\sum_{l=1}^{t-j}\binom{2n-2j}{l}\right)$, then for any element $v\in A$, we can find a unique pair of $i_1,i_2,\cdots,i_m$ and $j_1,j_2,\cdots,j_p$ such that $v^T=\sum_{a=1}^{m}H_X^{i_a}+\sum_{b=1}^{p}H_Z^{j_b}$. So the error mapped to the error syndrome $v$ is $E=\varphi'\left(\left(\{j_1,j_2,\cdots,j_p\}|\{i_1,i_2,\cdots,i_m\}\right)\right)$. Because the set $A$ contains all the vectors that corresponding to pairs of $i_1,i_2,\cdots,i_m$ and $j_1,j_2,\cdots,j_p$ which satisfy $1\leq w_s\left(\left(\{j_1,j_2,\cdots,j_m\}|\{i_1,i_2,\cdots,i_m\}\right)\right)\leq t$. Thereby, the collection of the errors $E$ that corresponding to pairs of $i_1,i_2,\cdots,i_m$ and $j_1,j_2,\cdots,j_p$ is $\overline{G}_{n|t}-\{I\}$. Finally, $C$ is a nondegenerate code. □

By theorem 2, we have the following corollary.

*Corollary 2.* Let $H$ be a check matrix such that any $4t$ columns are linearly independent, but there exists a set of $4t$ linearly dependent columns, then the stabilizer code $C(S)$ defined by $H$ has the minimum distance $d\geq 2t+1$. If $C(S)$ is nondegenerate, then its minimum distance satisfies $2t+1\leq d\leq 4t+1$.

*Proof*: The minimum distance of $C(S)$ is the quantum weight of the element in $N(S)$-$S$ with the minimum quantum weight, where $N(S)$ is the normalizer of $S$. Without lost of generality, suppose the $i_1$th,$i_2$th,$\cdots$,$i_m$th columns of $H_X$ and the $j_1$th,$j_2$th,$\cdots$,$j_p$th columns of $H_Z$ are linearly dependent, i.e., $v_{m,p}=\sum_{a=1}^{m}H_X^{i_a}+\sum_{b=1}^{p}H_Z^{j_b}=0$. Then, we have $m,p\geq 0$ and $m+p\geq 4t+1$. Let us assume as before that $E=\varphi'\left(\left(\{j_1,j_2,\cdots,j_p\}|\{i_1,i_2,\cdots,i_m\}\right)\right)$, then $E\in N(S)$ and

$$w_Q(E)=w_s\left(\left(\{j_1,j_2,\cdots,j_p\}|\{i_1,i_2,\cdots,i_m\}\right)\right)\geq 2t+1$$

Let us consider the case with $m+p=4t+1$:

Case (i): If $\left|\{j_1,j_2,\cdots,j_p\}\cap\{i_1,i_2,\cdots,i_m\}\right|=2t$, then $w_Q(E)=2t+1$;

if $\left|\{j_1,j_2,\cdots,j_p\}\cap\{i_1,i_2,\cdots,i_m\}\right|<2t$, then $w_Q(E)>2t+1$.

Case (ii): If $\left|\{i_1,i_2,\cdots,i_m\}\cap\{j_1,j_2,\cdots,j_p\}\right|=0$, then $w_Q(E)=4t+1$;

if $\left|\{i_1,i_2,\cdots,i_m\}\cap\{j_1,j_2,\cdots,j_p\}\right|>0$, then $w_Q(E)<4t+1$.

It follows from case (i) that $w_Q(E) > 2t+1$. Then we have $d \geq 2t+1$.

If $C(S)$ is nondegenerate, then
$$\min\{w_Q(e) \mid e \in N(S) - S\} \leq \min\{w_Q(e') \mid e \in S\}.$$

From which we see that $d \leq 4t+1$. Combining case (i) and case (ii), we get $2t+1 \leq d \leq 4t+1$. □

It is easy to see that some sequence of elementary row transformations and column permutation will transform the check matrix $H$ into the following standard form [3]:

$$R = \begin{array}{c} r\{ \\ n-r-k\{ \end{array} \begin{bmatrix} \overset{r}{I} & \overset{n-k-r}{A_1} & \overset{k}{A_2} & \overset{r}{B} & \overset{n-k-r}{0} & \overset{k}{C} \\ 0 & 0 & 0 & D & I & E \end{bmatrix} = [R_X \mid R_Z]$$

From which we see that the stabilizer code $C_R$ with check matrix $R$ is isomorphic to the stabilizer code $C_H = [[n,k,2t+1]]$ with check matrix $H$. Therefore, $C_H$ is nondegenerate if (and only if) $C_R$ is nondegenerate. Thus, to demonstrate that $C_R$ is nondegenerate it only has to show that $C_H$ is nondegenerate.

*Theorem 4.* If $r = n - k$ and there exists a column of $B$ with Hamming weight less than or equal to $t$-1, then $C_R$ is degenerate.

*Proof:* Follows from theorem 2. □

CSS codes are a special kind of stabilizer codes with the check matrix of the following form:

$$H = [H_X \mid H_Z] = \begin{bmatrix} A & 0 \\ 0 & B \end{bmatrix}$$

It is easy to see that, none of the columns of $H_X$ can be expressed as a linear combination of the columns of $H_Z$. Therefore, by theorem 2, we have the following corollary.

*Corollary 3.* The CSS code $C = [[n,k,2t+1]]$ is nondegenerate if and only if any $2t$ columns of $H_X$ and $H_Z$ are linearly independent respectively.

*Proof:* The condition is clearly sufficient. Then we will prove that it is necessary. Without lost of generality, suppose the $i_1$th, $i_2$th, $\cdots$, $i_t$th, $j_1$th, $j_2$th, $\cdots$, $j_t$th columns of $H_X$ are linearly dependent, then we have $\sum_{a=1}^{t} H_X^{i_a} + \sum_{b=1}^{t} H_X^{j_b} = 0$. Let the errors $E_1$ and $E_2$ be define as $E_1 = \prod_{k=1}^{t} Z_{i_k}$ and $E_2 = \prod_{k=1}^{t} Z_{j_k}$ respectively. Thus, $w_Q(E_1) = w_Q(E_2) = t$. It is easy to see that $s_{E_1}^T = \sum_{a=1}^{t} H_X^{i_a}$ and $s_{E_2}^T = \sum_{b=1}^{t} H_X^{j_b}$. Thus, $E_1$ and $E_2$ have the same error syndrome. Therefore, we produce a contradiction. □

*Corollary 4.* Let $H$ be a check matrix such that any $2t$ columns are linearly independent, but there exist a set of $2t$ linearly dependent columns, then the CSS code $C$ defined by $H$ has the minimum distance $d \geq 2t+1$. If $C$ is nondegenerate, then its minimum distance is $d = 2t+1$.

*Proof:* Our corollary follows from theorem 2 and corollary 3. □

The above assertion is similar to the classical case that the minimum distance of a classical linear code is determined by its parity check matrix, since that the CSS codes detect and correct quantum errors by making use of the error-correcting properties of the classical codes.

We know from corollary 2 and corollary 4 that the degenerate quantum stabilizer codes outperform the nondegenerate quantum stabilizer codes which had proved in reference [5].

## VI. CONCLUSION

We have presented some necessary and/or sufficient conditions for a stabilizer code over GF(2) is degenerate or nondegenerate for Pauli channel. It would be interesting to investigate the necessary and/or sufficient conditions for a stabilizer code over GF($q$) is degenerate or nondegenerate. It would be also interesting to investigate the necessary and/or sufficient conditions for a stabilizer code over GF($q$) is degenerate or nondegenerate for other quantum channels. It already proved that all CSS codes with alphabet size $q>4$, where $q$ is a prime power, must obey the Hamming bound [13]. Some special cases of interest are stabilizer codes and CSS codes over a nonprime power alphabet and small alphabet $q < 5$. We hope our work will helpful for solving this problem and for design quantum degenerate codes with good performance.